\documentclass[%
 reprint,
 amsmath,amssymb,
 aps,
floatfix,
longbibliography,
]{revtex4-1}

\usepackage{graphicx}
\usepackage{dcolumn}
\usepackage{bm}

\usepackage{multirow}
\usepackage{array}

\begin{document}

\preprint{APS/123-QED}

\title{Dual-polarized all-angle cloaking of a dielectric nanowire by helical graphene ribbons}

\author{Vitalii I. Shcherbinin$^{1,2}$}
\author{Yuliya K. Moskvitina$^{2}$}
\author{Volodymyr I. Fesenko$^{1,3}$}  
 \email{volodymyr.i.fesenko@gmail.com}
\author{Vladimir~R.~Tuz$^{1,3}$}
\affiliation{$^1$International Center of Future Science, State Key Laboratory of Integrated Optoelectronics, College of Electronic Science and Engineering, Jilin University, 2699 Qianjin St., Changchun 130012, China}
\affiliation{$^2$National Science Center `Kharkiv Institute of Physics and Technology' of National Academy of Sciences of Ukraine, 1, Akademicheskaya Str., Kharkiv 61108, Ukraine}
\affiliation{$^3$Institute of Radio Astronomy of National Academy of Sciences of Ukraine, 4 Mystetstv St., Kharkiv 61002, Ukraine}

\date{\today}

\begin{abstract}
Scattering from a dielectric nanowire coated by helical graphene strips (nanoribbons) is investigated for dual-polarized wave at oblique incidence. In the long-wavelength approximation, the helical strips are treated as a homogeneous layer with averaged tensor conductivity. It is shown that performance of well-known surface cloaks in the form of graphene monolayer, axial and azimuthal graphene strips can be deteriorated in a wide range of incidence angles. To overcome this problem, helical graphene strips are proposed as an advanced metasurface for dual-polarized all-angle cloaking of dielectric nanowire in the terahertz band. It is found that such metasurface suppresses scattering from nanowire more effectively as compared to graphene monolayer, regardless of the angle of wave incidence. Moreover, dual-polarized all-angle cloaking of dielectric nanowire can be broadly tuned in frequency with parameters of helical graphene strips.
\end{abstract}



\maketitle
\section{\label{intro}Introduction}
Electromagnetic cloaking of objects, which makes them less visible to observer, is a subject of much interest for decades \cite{Fleury_2015}. At present, there is a number of well-established cloaking methods, ranging from transformation-based cloaking for microwaves to plasmonic one at infrared and visible frequencies. Our interest is in mantle cloaking of dielectric objects by conducting surfaces \cite{Alu_2009}. Such a method relies on scattering cancellation due to destructive interference between fields scattered from the object and the cloak \cite{Alu_2005}. One of its advantages is thinness of the cloak, which, in contrast to transformation-based and plasmonic counterparts, can be conformal to the object. Another advantage is that the mantle cloak does not isolate the object electromagnetically from ambient space. Together, these beneficial properties make mantle cloaking particularly suitable for use in novel sensor and antenna applications \cite{Alu_2010, Vellucci_2017}. In these applications, objects to be cloaked usually have a cylindrical form.

In the THz frequency band, graphene is known to have a number of unique properties to serve as a thinnest possible mantle cloak \cite{Chen_2011}. First, graphene is good conducting at THz frequencies. Second, its intrinsic losses are low enough to make cloaking performance of graphene monolayer close to that of an ideal cloak. Third, graphene conductivity can be tuned by applied bias voltage, thereby providing possibility for dynamical tuning of cloaking frequency. All these properties ensure effective scattering suppression for single and multiple dielectric cylinders with uniform graphene coating \cite{Chen_2011, Riso_2015, Naserpour_2017, Fesenko_2018}. Cloaking performance can be further improved using periodic array of sub-wavelength graphene nanopatches \cite{Chen_2013}, which act as metasurface with averaged conductivity. Such a cloak demonstrates good performance for both dielectric and metallic cylindrical objects, as well as for elliptical structures \cite{Bernety_2015} and conducting wedges \cite{Forouzmand_2015}. Along with graphene nanopatches, other sub-wavelength elements find use in design of graphene-based metasurfaces. Example is a metasurface composed of periodic graphene nanodisks \cite{Danaeifar_2016}, which make it possible to enhance frequency bandwidth of cylindrical mantle cloak. It should be emphasized that good cloaking performance of graphene monolayer and graphene-based metasurfaces is demonstrated in \cite{Chen_2011, Chen_2013, Riso_2015, Bernety_2015, Forouzmand_2015, Naserpour_2017, Fesenko_2018, Danaeifar_2016} for objects illuminated by normally incident waves of single (TE or TM) polarization.

Ideally, mantle cloak should maximally reduce wave scattering from a given object in a broad frequency band, as well as be insensitive to polarization of incident wave and angle of incidence. As emphasized in \cite{Alu_2010_2}, design of such an advanced cloak is a challenging task, mainly because of the cross-polarization coupling. Therefore, to our best knowledge, the mantle cloak capable of meeting all above-mentioned requirements is not yet available. For dielectric and metallic objects in microwave frequency band, an anisotropic metasurface cloak, which can minimize scattering of both TE- and TM-polarized incident waves with the same frequencies, is investigated in \cite{Monti_2015}. The metasurface of interest is in the form of an array of orthogonal metallic strips. In the case of dielectric objects, its cloaking principle relies on selective ability of conducting strips to suppress scattering of electromagnetic waves having electric field parallel to the direction of strips \cite{Padooru_2012}. Thus, the widths of axial and azimuthal strips around a dielectric cylinder can be varied to manipulate scattering of waves with orthogonal TM and TE polarizations independently and, if necessary, to achieve simultaneous suppression of their scattering at desired cloaking frequency. In \cite{Monti_2015}, metasurface composed of metallic strips was designed to suppress scattering of both TM- and TE-polarized waves from a given object under normal incidence. Therefore, in the case of oblique illumination, which is always associated with cross coupling of TE and TM polarizations for scattered field, cloaking performance of this metasurface can be nonoptimal. 

Hence, to achieve dual-polarized all-angle cloaking of a dielectric nanowire in the THz band, it is necessary to apply a new design of a graphene-based metasturface. In such a design, it is important to keep in mind that there is a fundamental trade-off between bandwidth and total scattering reduction provided by a cloak \cite{Fleury_2015, Monticone_2016}. Therefore, our investigation is not concerned with the requirement on frequency bandwidth. As a cloaking metasurface, we consider helical graphene strips (ribbons). In the long-wavelength approximation, such strips are characterized by the averaged conductivity of tensor form \cite{Shcherbinin_2018}. At present, there are a lot of fabrication techniques of graphene nanoribbons \cite{Xu_2016}. This fact inspires the hope that one of these techniques can be useful in fabrication of helical graphene strips under consideration.

\section{\label{Scattering} Nanowire coated by helical graphene strips}

\begin{figure}[t!]
\centering
\includegraphics[width=0.4\linewidth]{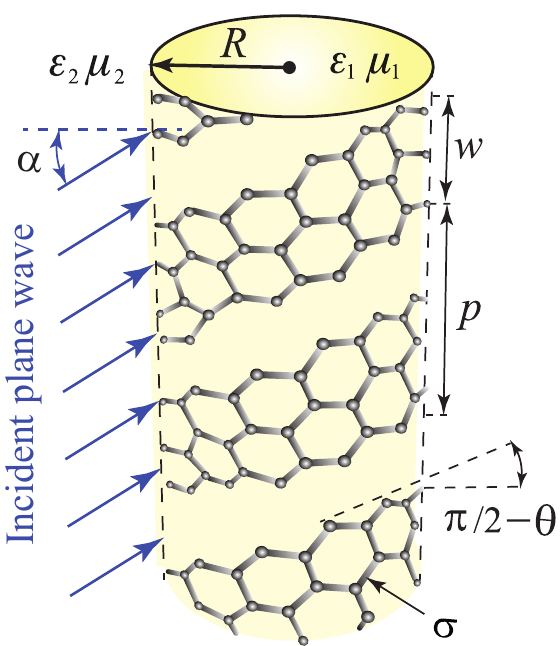}
\caption{Dielectric cylinder coated by helical graphene strips for oblique wave incidence.}
\label{fig:sketch}
\end{figure}

Consider a cylindrical wire of radius $R$, relative permittivity $\varepsilon_1$ and permeability $\mu_1$. The wire is coated by helical graphene strips and embedded in outside ambient with relative constitutive parameters $\varepsilon_2$ and $\mu_2$. Graphene strips have the pitch angle $\theta$, the width $w$ and the period $p$ (Fig.~\ref{fig:sketch}). 

The graphene-coated wire is under oblique incidence of the plane wave with the frequency $\omega$. In cylindrical coordinates $\{r,\varphi,z\}$, the incident field is considered to have the general form \cite{Bohren_1998}
\begin{equation}
\left\{ {\begin{matrix}
   {H_z^i} \cr
   {E_z^i} \cr
\end{matrix}
} \right\} = \left\{ {\begin{matrix}
   {H_0} \cr
   {E_0} \cr
\end{matrix}
} \right\}\sum_{n=-\infty}^\infty i^n J_n(k_2r)F_n, \label{eq:eq1}
\end{equation}
where $H_0$ and $E_0=PH_0$ are the field amplitudes, $k_2^2=\varepsilon_2\mu_2k^2-k_z^2$, $k=\omega/c$, $k_z=k\sin\alpha$, $\alpha$ is the incidence angle with respect to the $\{r,\varphi\}$-plane, $J_n(\cdot)$ is the Bessel function of $n$-th order, $F_n=\exp(-i\omega t + ik_zz + in\varphi)$. For $E_0=0$ ($P=0$) and $H_0=0$ ($P=\infty$) the illuminating wave has TE and TM polarizations, respectively, while for $|P|=1$ the wave is dual-polarized. 

The scattered field and the field inside the wire can be expanded in terms of azimuthal harmonics as 
\begin{equation}
\left\{ {\begin{matrix}
   {H_z^s} \cr
   {E_z^s} \cr
\end{matrix}
} \right\} = H_0\sum_{n=-\infty}^\infty\left\{ {\begin{matrix}
   {c_n} \cr
   {d_n} \cr
\end{matrix}
} \right\} i^n H_n^{(1)}(k_2r)F_n, \label{eq:eq2}
\end{equation}
\begin{equation}
\begin{split}
\left\{ {\begin{matrix}
   {H_{z1}} \cr
   {E_{z1}} \cr
\end{matrix}
} \right\} &= H_0\sum_{n=-\infty}^\infty\left\{ {\begin{matrix}
   {a_n} \cr
   {b_n} \cr
\end{matrix}
} \right\} i^n J_n(k_1r)F_n \\
&=\sum_{n=-\infty}^\infty\left\{ {\begin{matrix}
   {H_{z1}^n} \cr
   {E_{z1}^n} \cr
\end{matrix}
} \right\}F_n, \label{eq:eq3}
\end{split}
\end{equation}
respectively. Here $k_1^2=\varepsilon_1\mu_1k^2-k_z^2$, $H_n^{(1)}(\cdot)$ is the Hankel function of the first kind, $a_n$, $b_n$, $c_n$, and $d_n$ are the unknowns, which are found from the conditions imposed on the wave field at the wire surface $r=R$. 

Assuming that the wavelength $\lambda=2\pi/k$ of the incident wave far exceeds the period $p$ of graphene strips, one can approximate the strip-loaded surface of the wire by a homogeneous surface (graphene-based metasurface). The resulting metasurface possesses tensor conductivity and is governed by the following averaged boundary conditions at $r=R$ (see Ref.~\onlinecite{Shcherbinin_2018} and Appendix A):
\begin{equation}
\begin{split}
& E_{z1}^n = E_{z2}^n,~ H_{z2}^n - H_{z1}^n = -\sigma_{\varphi\varphi} E_{\varphi 1}^n - \sigma_{\varphi z} E_{z1}^n, \\
& E_{\varphi 1}^n = E_{\varphi 2}^n,~ H_{\varphi 2}^n - H_{\varphi 1}^n =\sigma_{zz} E_{z 1}^n + \sigma_{z\varphi} E_{\varphi 1}^n,
\label{eq:eq4} 
\end{split}
\end{equation}
where $\sigma_{\varphi\varphi} = \sigma_\perp\cos^2\theta + \sigma_\parallel \sin^2\theta$, $\sigma_{zz} = \sigma_\parallel \cos^2\theta + \sigma_\perp\sin^2\theta$, $\sigma_{\varphi z} = \sigma_{z\varphi} = (\sigma_\parallel - \sigma_\perp)\sin\theta\cos\theta$, $\sigma_{\perp,\parallel} = (Z_{\perp,\parallel} + Z_g^{av})^{-1}$, $Z_g^{av}=u/\sigma$, $u=w/p$, $\sigma=\sigma(\omega)$ is the conductivity of graphene strips,
\begin{equation}
\begin{split}
& Z_\parallel = -iqZ_0N\mu_{\alpha}\ln[\sin^{-1}(0.5\pi u)], \\
& Z_\perp = iZ_0\left\{qN\varepsilon_{\alpha}\ln[\sin^{-1}(0.5\pi(1- u))]\right\}^{-1},
\label{eq:eq5} 
\end{split}
\end{equation}
and $q=1-k_\parallel^2/(\varepsilon_\alpha\mu_\alpha k^2)$, $N=2p/\lambda$, $\varepsilon_{\alpha} = \varepsilon_{1} + \varepsilon_{2}$, $\mu_{\alpha}=\mu_{1}\mu_{2}/(\mu_{1} + \mu_{2})$, $Z_0$ is the impedance of free space, $k_\parallel^2=(k_z\sin\theta + nR^{-1}\cos\theta)^2$. 

In Eq.~(\ref{eq:eq4}), 
\begin{equation}
\left\{ {\begin{matrix}
   {H_{z2}} \cr
   {E_{z2}} \cr
\end{matrix}
} \right\} = 
\left\{ {\begin{matrix}
   {H_{z}^i+H_{z}^s} \cr
   {E_{z}^i+E_{z}^s} \cr
\end{matrix}
} \right\} =
\sum_{n=-\infty}^\infty\left\{ {\begin{matrix}
   {H_{z2}^n} \cr
   {E_{z2}^n} \cr
\end{matrix}
} \right\} F_n, \label{eq:eq6}
\end{equation}
and $\varphi$-components of the electromagnetic field are expressed in terms of Eqs.~(\ref{eq:eq1})-(\ref{eq:eq3}). It should be noted that the boundary conditions (\ref{eq:eq4}) are accurate to at least first-order in $p/\lambda$ \cite{Tretyakov_2003} and are valid for waves propagating in an arbitrary direction with respect to the strips.

For axial ($\theta=0^\circ$) and azimuthal ($\theta=90^\circ$) graphene strips one has $\sigma_{z \varphi} = \sigma_{\varphi z} = 0$. In this case, assuming that the strip conductivity is high enough ($Z_g^{av}\approx 0$), one can easily reduce Eqs.~(\ref{eq:eq1})-(\ref{eq:eq4}) to the problem of wave scattering from a dielectric cylinder with periodic vertical and circumferential metallic strips for normal ($k_z=0$) \cite{Padooru_2012} or oblique \cite{Sipus_2018} incidence. For these structures reasonable accuracy of the averaged boundary conditions (\ref{eq:eq4}) can be shown by comparison with full-wave simulations \cite{Padooru_2012, Sipus_2018}. 

Substitution of Eqs.~(\ref{eq:eq1})-(\ref{eq:eq3}) into Eq. (\ref{eq:eq4}) gives the system of equations for unknown amplitudes $a_n$, $b_n$, $c_n$, and $d_n$ of azimuthal spatial harmonics in relation to the cross-polarization parameter $P$ of the incident wave. With the amplitudes $c_n$ and $d_n$ of scattered field, one obtains the scattering efficiency \cite{Bohren_1998}
\begin{equation}
Q_\textrm{sca} = \frac{2}{kR(1+|P|^2)}\sum_{n=-\infty}^\infty \left(|c_n|^2+|d_n|^2\right),
\label{eq:eq7} 
\end{equation}
which serves as a measure of visibility for a cylindrical wire at given wave frequency and angle of incidence. Note that in the case of metasurface composed of helical graphene strips, the cross-polarization coupling cannot be neglected for scattered wave, even though normally incident wave is considered. This is because the metasurface conductivity has a form of tensor with non-zero off-diagonal components $\sigma_{z\varphi}$ and $\sigma_{\varphi z}$. Such a property distinguishes helical graphene strips from other metasurfaces aimed at electromagnetic cloaking of cylindrical objects. 

\section{\label{Cloaking}Cloaking with graphene monolayer, axial and azimuthal graphene strips}

\begin{figure}[t!]
\centering
\includegraphics[width=0.9\linewidth]{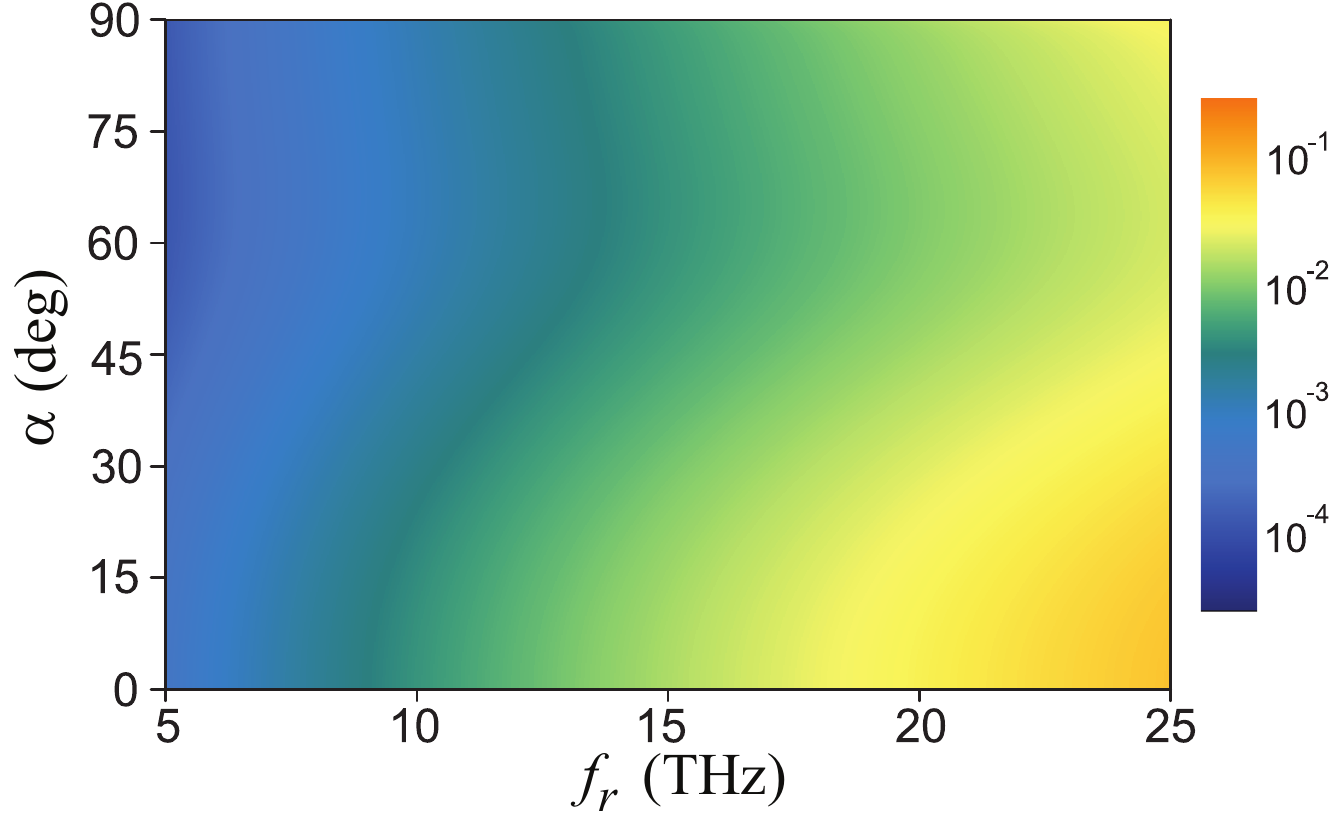}
\caption{Scattering efficiency $Q_\textrm{sca}$ of a bare dielectric nanowire illuminated by dual-polarized wave at oblique incidence.}
\label{fig:fig2}
\end{figure} 

As the object to be cloaked, we consider a dielectric nanowire in free space ($\varepsilon_2=1$, $\mu_2=1$). The nanowire is under oblique incidence of the dual-polarized plane wave and has the following parameters: $R=0.5$~$\mu$m, $\varepsilon_1=3.9$ (SiO$_2$) and $\mu_1=1$. For normal incidence ($\alpha=0$) of TE-polarized ($P=0$) and TM-polarized ($P=\infty$) waves, scattering characteristics of such nanowire coated by a graphene monolayer ($u=1$) can be found in \cite{Riso_2015, Naserpour_2017, Fesenko_2018} and are used as benchmark for our simulations. Note that these characteristics have been validated against those obtained by the COMSOL Multiphysics\textsuperscript{\textregistered} software \cite{Naserpour_2017}. In the following, the conductivity $\sigma(\omega)$ of graphene coating, if any, is found from the widely-accepted Kubo formula \cite{Falkovsky_2007} for the temperature of 300~K, the chemical potential of 0.5~eV, and the relaxation time $1.84\times 10^{-13}$ s (see Fig. 1(b) of Ref. \onlinecite{Yu_2018} and Appendix B).

For reference, we first investigate the visibility of a bare ($u=0$) dielectric nanowire under oblique incidence of the dual-polarized plane wave. Fig.~\ref{fig:fig2} shows scattering efficiency $Q_\textrm{sca}$ of the dielectric nanowire as a function of the wave frequency $f_r=\omega/(2\pi)$ and the incidence angle $\alpha$. It can be seen that $Q_\textrm{sca}$ increases with increasing frequency of illuminating wave for any $\alpha$ \cite{Chen_2011, Chen_2013, Soric_2013, Riso_2015}. Generally, the higher is the incidence angle, the lower is the scattering from the nanowire. The exceptional angles are those near the grazing incidence ($\alpha\approx 90^\circ$), which corresponds to increased contribution of TE-polarization to scattered field due to cross-polarization coupling. 

\begin{figure}[t!]
\centering
\includegraphics[width=0.8\linewidth]{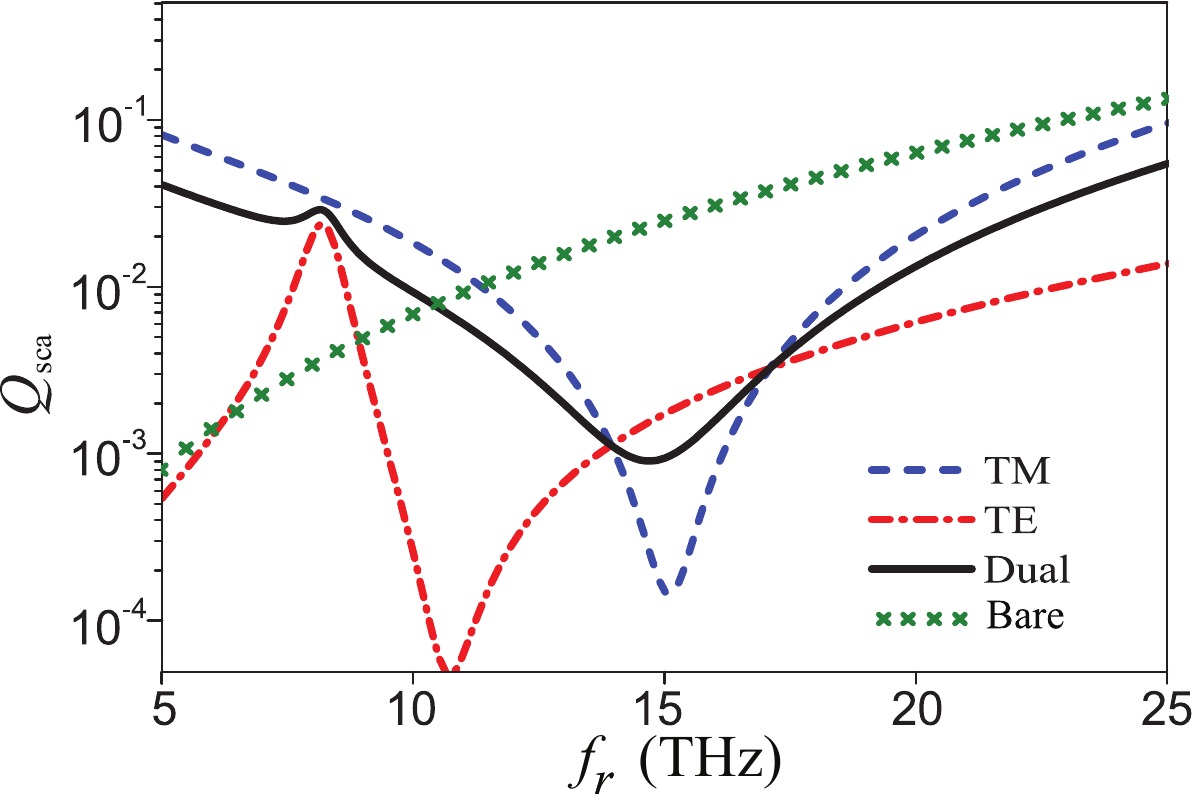}
\caption{Scattering efficiency versus wave frequency for a graphene-coated nanowire under normal incidence ($\alpha=0^\circ$) of TE, TM and dual-polarized waves. Scattering efficiency of bare nanowire illuminated by normally incident dual-polarized wave is shown for comparison}
\label{fig:fig3}
\end{figure}

\begin{figure}[t!]
\centering
\includegraphics[width=0.9\linewidth]{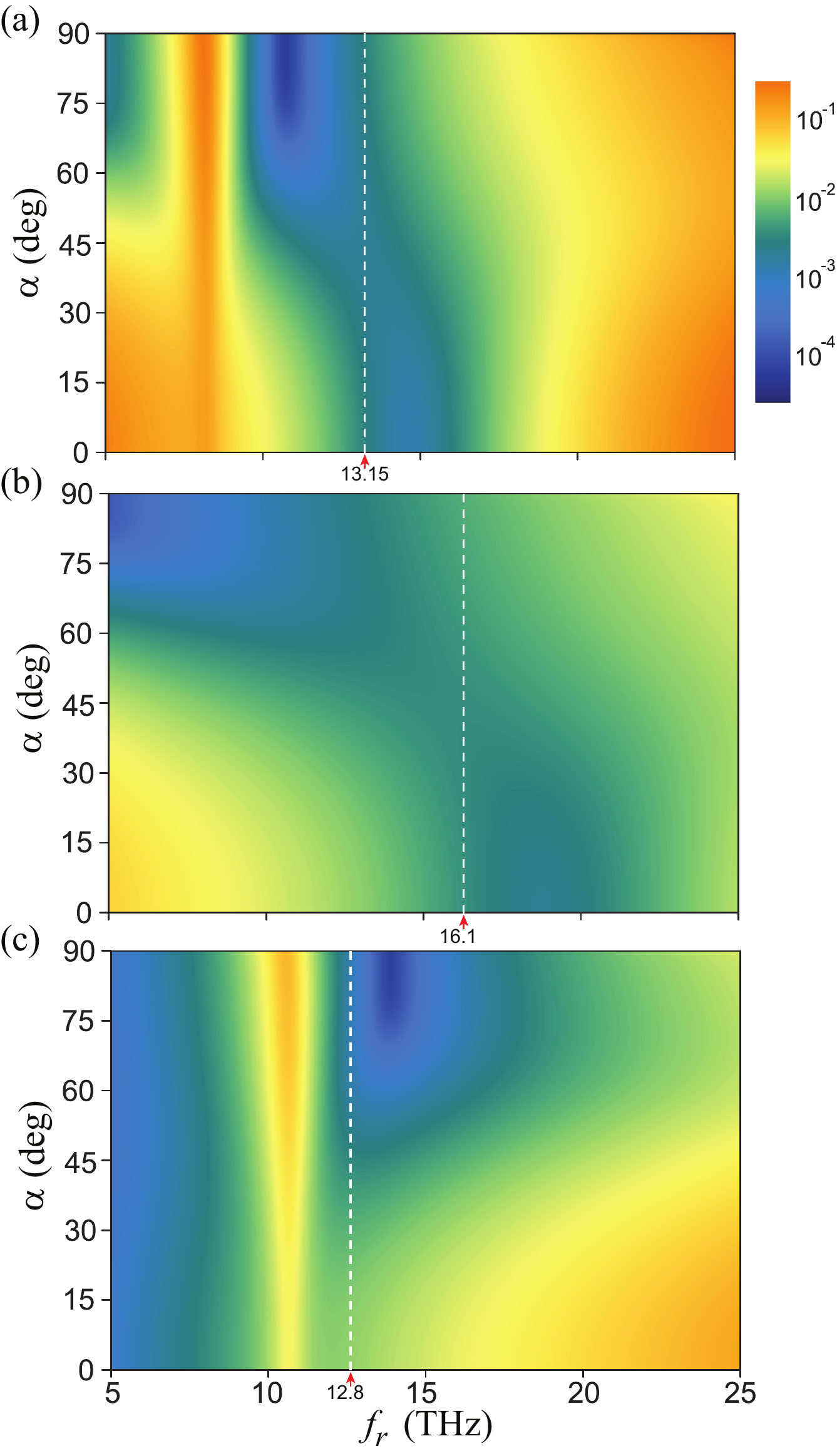}
\caption{Scattering efficiency $Q_\textrm{sca}$ versus wave frequency and incidence angle for a dielectric nanowire coated by: (a) graphene monolayer, (b) axial and (c) azimuthal graphene strips at oblique incidence of dual-polarized wave. In the cases (a), (b) and (c), the optimal frequencies of all-angle cloaking equal 13.15~THz ($Q_\textrm{sca}^\textrm{max}=1.8\times 10^{-3}$), 16.1~THz $(Q_\textrm{sca}^\textrm{max}=1.3\times 10^{-2})$ and 12.8~THz $(Q_\textrm{sca}^\textrm{max}=2.1\times 10^{-2})$, respectively, and are shown by the white dashed lines.
}
\label{fig:fig4}
\end{figure}

It is well-known that homogeneous graphene coating ($u=1$) can drastically suppress the scattering from a dielectric cylinder in certain frequency range \cite{Chen_2011, Riso_2015}. For normally incident TE, TM and dual-polarized waves this effect can be clearly seen from Fig.~\ref{fig:fig3}. Note that for scattered field there is no cross-polarization coupling in this case. As Fig.~\ref{fig:fig3} suggests, for TE- and TM-polarized incident waves the minimum scattering from graphene-coated nanowire is attained at different frequencies \cite{Naserpour_2017}. As a consequence, when compared to these waves, for normally incident dual-polarized wave the minimum scattering efficiency is shifted in frequency and increases at least by an order of magnitude. The cloaking ability of the graphene monolayer is further deteriorated with increasing incidence angle $\alpha$ [Fig.~\ref{fig:fig4}(a)]. In this process, the minimum scattering efficiency shifts towards lower frequencies. As mentioned above, this is explained by increase in TE-polarized part of scattered field \cite{Alu_2010}.

As can be seen from Fig.~\ref{fig:fig4}(a), for any $\alpha$ the dielectric nanowire can be effectively cloaked by the graphene monolayer in certain frequency range of dual-polarized incident wave, but features enhanced scattering in other frequencies. In this regard, the most optimal frequency of all-angle cloaking is 13.15~THz, which corresponds to relatively low visibility of the wire for any incidence angle. In this case, the maximal scattering efficiency $Q_\textrm{sca}^\textrm{max}=\text{max} \{Q_\textrm{sca}(\alpha)\}$ is about $1.8\times 10^{-3}$. In principle, the optimal frequency of all-angle cloaking can be tuned with chemical potential of graphene. Such a possibility is a matter of common knowledge \cite{Chen_2011, Chen_2013, Riso_2015, Naserpour_2017, Fesenko_2018} and therefore is not considered in our study.

For generality, we also investigate the ability of extensively studied axial and azimuthal strips to cloak the dielectric nanowire under oblique incidence of dual-polarized wave. In all our subsequent simulations, the period of strips is kept constant and fulfills the requirement  $N=0.1kR$. In the case of $u=0.6$, Figs.~\ref{fig:fig4}(b) and \ref{fig:fig4}(c) show the scattering efficiency in \{$f_r,\alpha$\}-plane for a dielectric nanowire coated by axial ($\theta=0^\circ$) and azimuthal ($\theta=90^\circ$) graphene strips, respectively. Compared to the graphene monolayer [see Fig.~\ref{fig:fig4}(a)], such strips induce much larger scattering in a wide range of frequencies, including optimal ones for all-angle cloaking. In the case of axial and azimuthal graphene strips with $u=0.6$, these optimal frequencies equal 16.1~THz and 12.8~THz, respectively. It is evident that such a cloaking satisfies the condition $Q_\textrm{sca}^\textrm{max}>10^{-2}$ and therefore is far from ideal.

It is seen from Figs.~\ref{fig:fig4}(b) and \ref{fig:fig4}(c) that the axial strips provide the most effective scattering cancellation for normally incident wave, while azimuthal strips possess the improved ability to cloak the dielectric nanowire under grazing incidence. This is explained by two facts. First, conducting strips suppress scattering of waves, which have electric field parallel to the strips \cite{Padooru_2012}. Second, for normal incidence ($\alpha=0^\circ$) there is mostly TM-polarized scattered wave with dominant axial electric field. Contribution of TE-polarization to scattered field is minor in this case (see Fig.~\ref{fig:fig3}). This contribution grows in importance with increasing incidence angle and attains maximum at grazing incidence ($\alpha=90^\circ$). From the above reasoning, it is of interest to investigate the all-angle scattering characteristics of helical graphene strips coiled around dielectric nanowire to be cloaked.

\section{\label{angles}Dual-polarized all-angle cloaking with helical graphene strips}

\begin{figure}[htp]
\centering
\includegraphics[width=0.9\linewidth]{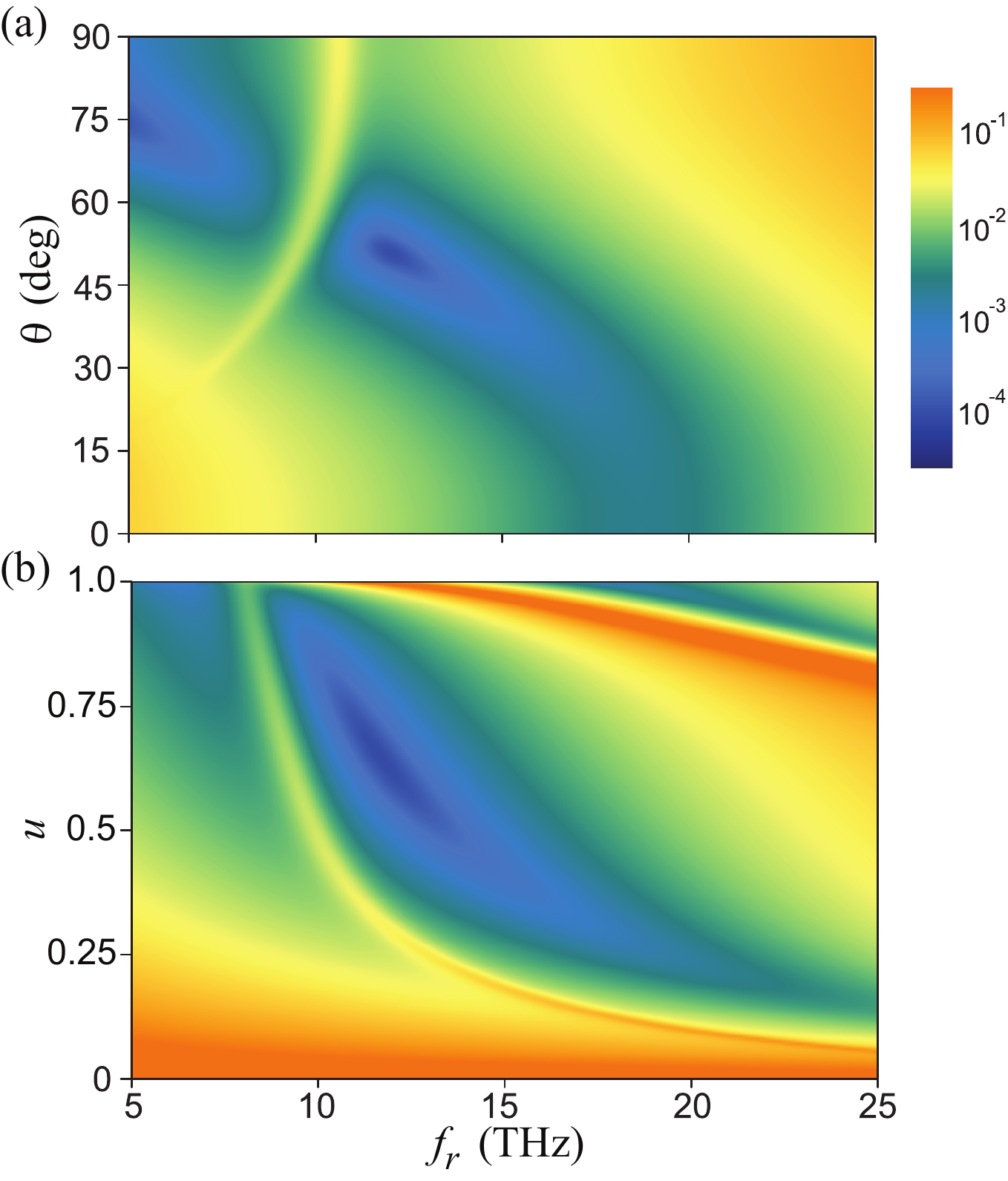}
\caption{Frequency-dependent scattering efficiency of a dielectric nanowire illuminated by normally incident ($\alpha=0^\circ$) dual-polarized wave as a function of (a) pitch angle $\theta$ ($u=0.6$) and (b) width-to-period ratio $u$ ($\theta=50^\circ$) of helical graphene strips.}
\label{fig:fig5}
\end{figure}

\begin{figure}[htp]
\centering
\includegraphics[width=0.8\linewidth]{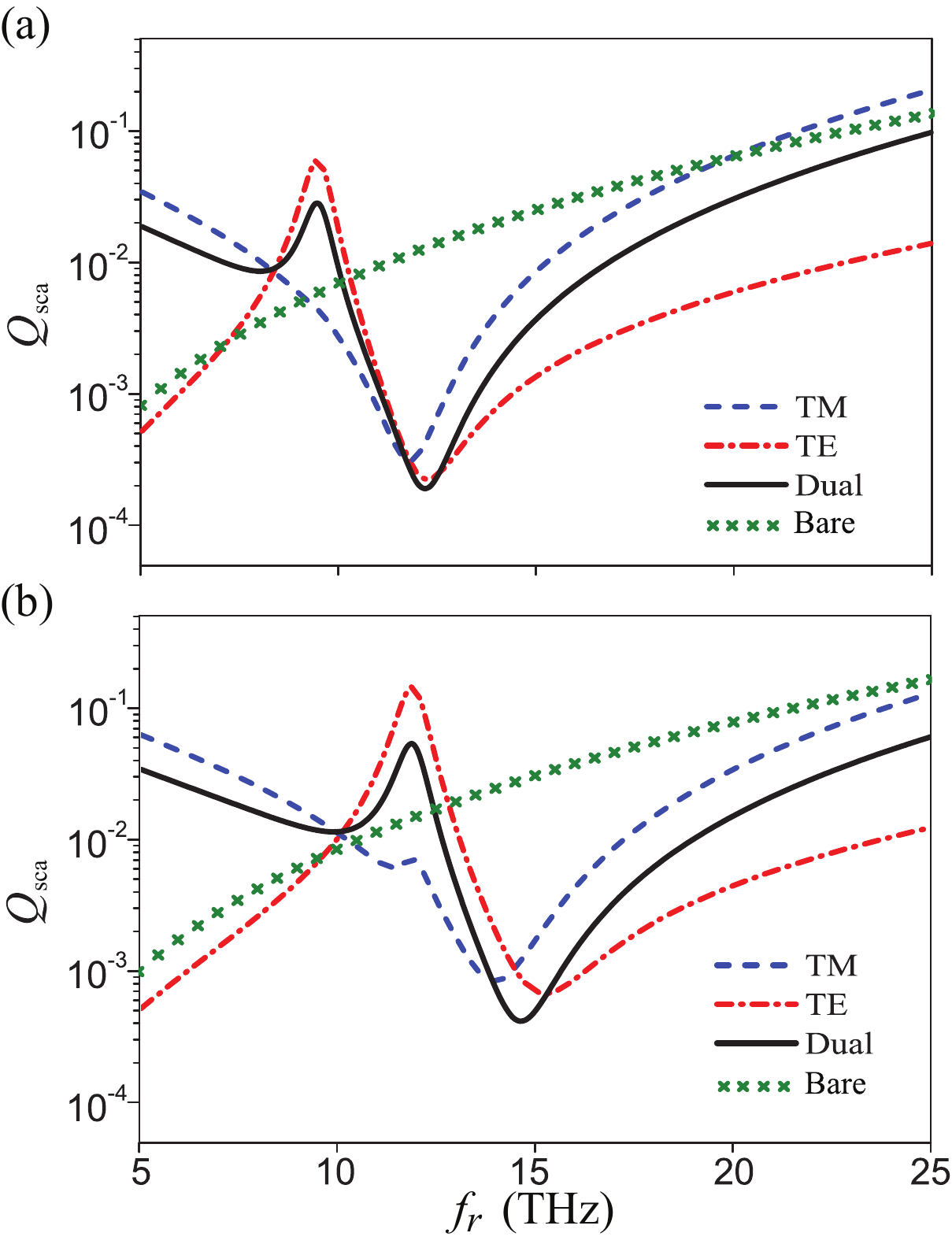}
\caption{The same as in Fig.~\ref{fig:fig3}, but for a dielectric nanowire coated by helical graphene strips with (a) $u=0.6$,  $\theta=50^\circ$ and (b) $u=0.36$,  $\theta=55^\circ$.}
\label{fig:fig6}
\end{figure}

First we consider dual-polarized wave normally incident ($\alpha=0^\circ$) on dielectric nanowire coated by graphene strips having the pitch angle $\theta$ and the width-to-period ratio $u$. For $u=0.6$ the scattering efficiency $Q_\textrm{sca}$ as a function of $\theta$ and frequency of illuminating wave is shown in Fig.~\ref{fig:fig5}(a). The minimum scattering shifts towards lower frequencies with increase in pitch angle. What is more interesting, in the case of helical graphene strips, which always lead to cross-polarization coupling for scattered field, the minimum scattering efficiency is lower than that for both axial and azimuthal strips. For any nonzero width of graphene strips there is an absolute scattering minimum relative to a certain value of the pitch angle $\theta$. In the case of $u=0.6$, this value is about $50^\circ$. Formation of such minimal visibility is due to the fact that helical graphene strips, as opposed to graphene monolayer (Fig.~\ref{fig:fig3}), provide effective scattering cancellation for TE- and TM-polarized incident waves with close frequencies (Fig.~\ref{fig:fig6}). This demonstrates beneficial effect of cross-polarization coupling induced by strips. 

For $\theta=50^\circ$ and variable values of $f_r$ and $u$ the scattering efficiency is depicted in Fig.~\ref{fig:fig5}(b). As is seen from this figure [compare also Figs.~\ref{fig:fig3} and \ref{fig:fig6}(b)], electromagnetic cloaking of dielectric nanowire by helical graphene strips can be more effective as compared to cloaking by a graphene monolayer, which is known to be highly efficient \cite{Chen_2011}. Note that there is no continuous transition of graphene-coated nanowire to bare dielectric cylinder as $u$ approaches zero in Eq.~(\ref{eq:eq5}). In the extreme case of thin strips, one comes to unidirectional conductor, which is conducting ($Z_\parallel$ is finite and approaches zero as $p\to 0$) in the direction parallel to the strips and perfectly insulating ($Z_\perp\to\infty$) in the perpendicular direction. Scattering problem for a dielectric cylinder coated by such helical strips made of perfect electric conductor has been studied extensively in Refs.~\onlinecite{Hady_2008, Chen_2011_2}.

\begin{figure}[htp]
\centering
\includegraphics[width=0.9\linewidth]{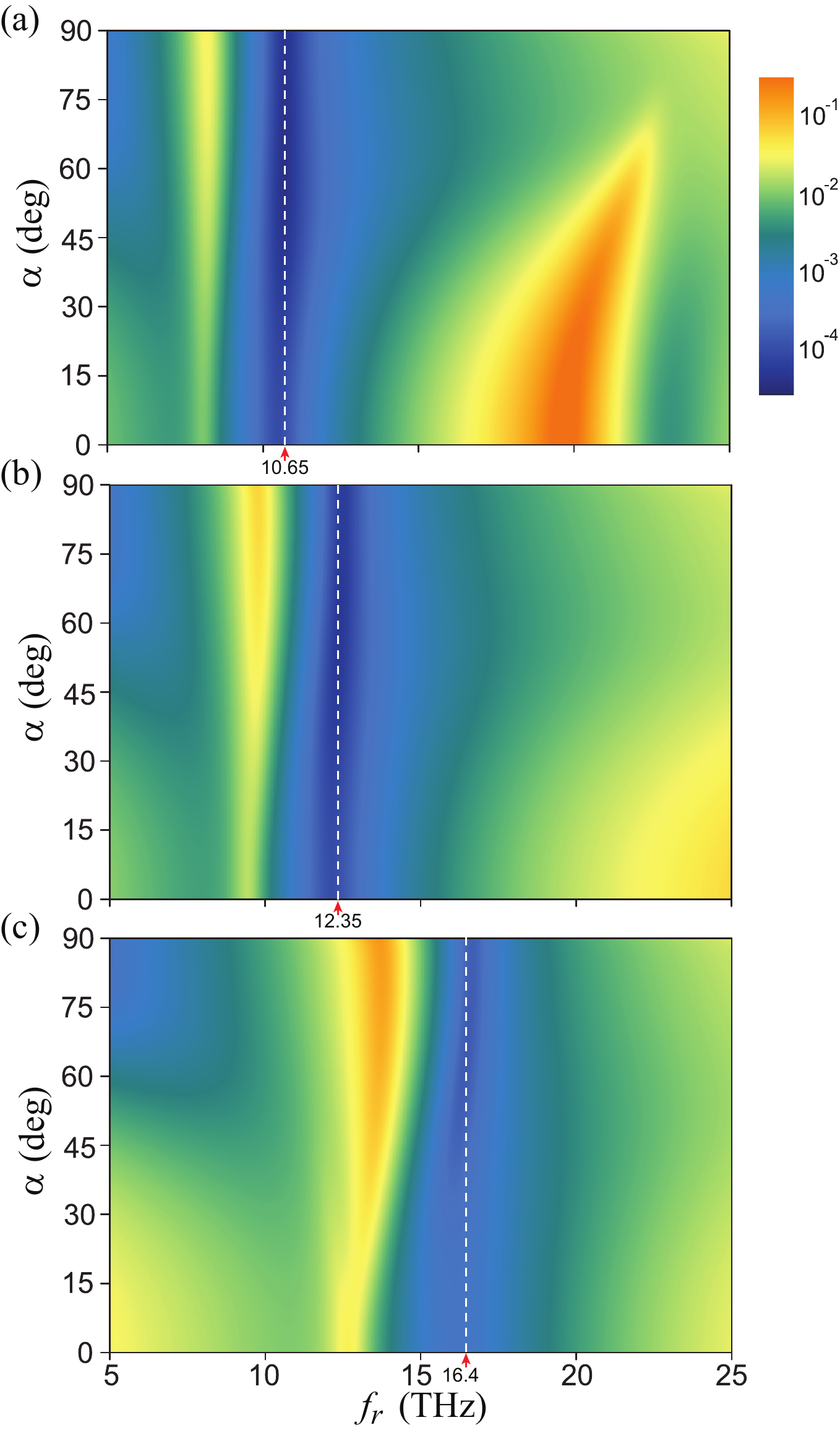}
\caption{Dual-polarized all-angle cloaking of a dielectric nanowire by helical graphene strips at frequencies (a) 10.65 THz ($u=0.9$,  $\theta=41^\circ$, $Q_\textrm{sca}^\textrm{max}=1.8\times 10^{-4}$), (b) 12.35 THz ($u=0.6$,  $\theta=50^\circ$, $Q_\textrm{sca}^\textrm{max}=2\times 10^{-4}$), and (c) 16.4 THz  ($u=0.3$,  $\theta=54^\circ$, $Q_\textrm{sca}^\textrm{max}=8.1\times 10^{-4}$).}
\label{fig:fig7}
\end{figure}

\begin{figure}[htp]
\centering
\includegraphics[width=0.9\linewidth]{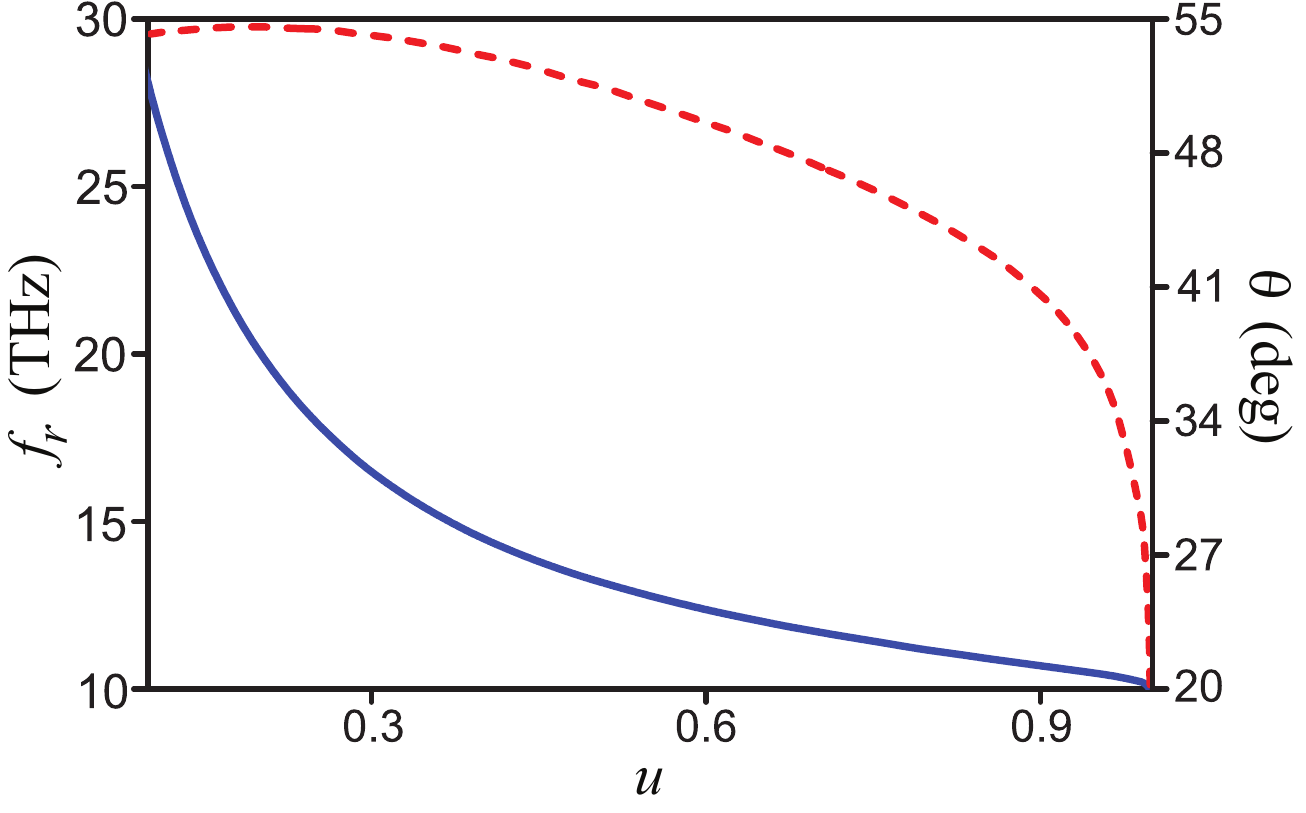}
\caption{Frequency $f_r$ and pitch angle $\theta$ versus the strip width for optimal dual-polarized all-angle cloaking of a dielectric nanowire by helical graphene strips.}
\label{fig:fig8}
\end{figure}

As Fig.~\ref{fig:fig5} suggests, at given angle of incidence $\alpha$ we can use parameters of graphene strips to tune the minimum visibility of dielectric nanowire and thus to achieve the most effective scattering suppression at desired frequency for some $u$ and $\theta$. By proceeding similarly for different $\alpha$ we obtain the optimal values of $u$, $\theta$ and frequency, which correspond to the lowest averaged scattering for all incidence angles. For $u=0.9$ ($\theta=41^\circ$), $u=0.6$ ($\theta=50^\circ$) and $u=0.3$ ($\theta=54^\circ$) the scattering efficiency $Q_\textrm{sca}$ as a function of frequency $f_r$ and incidence angle $\alpha$ of illuminating wave is depicted in Fig.~~\ref{fig:fig7} and shows the best performance for all-angle cloaking of dielectric nanowire at frequencies about 10.65~THz, 12.35~THz and 16.4~THz, respectively. As $u$ decreases, the cloaking ability of helical graphene strips is somewhat degraded. Despite this, in the frequency range of interest such strips suppress scattering from nanowire more effectively as compared to graphene monolayer [see Fig.~\ref{fig:fig4}(a)]. The wide tuning band of the cloak in the form of helical graphene strips can be also seen from Fig.~\ref{fig:fig8}, which shows the influence of strip width on optimal values of the pitch angle $\theta$ and the wave frequency $f_r$ for all-angle cloaking of dielectric nanowire.

\begin{figure}[htp]
\centering
\includegraphics[width=1.0\linewidth]{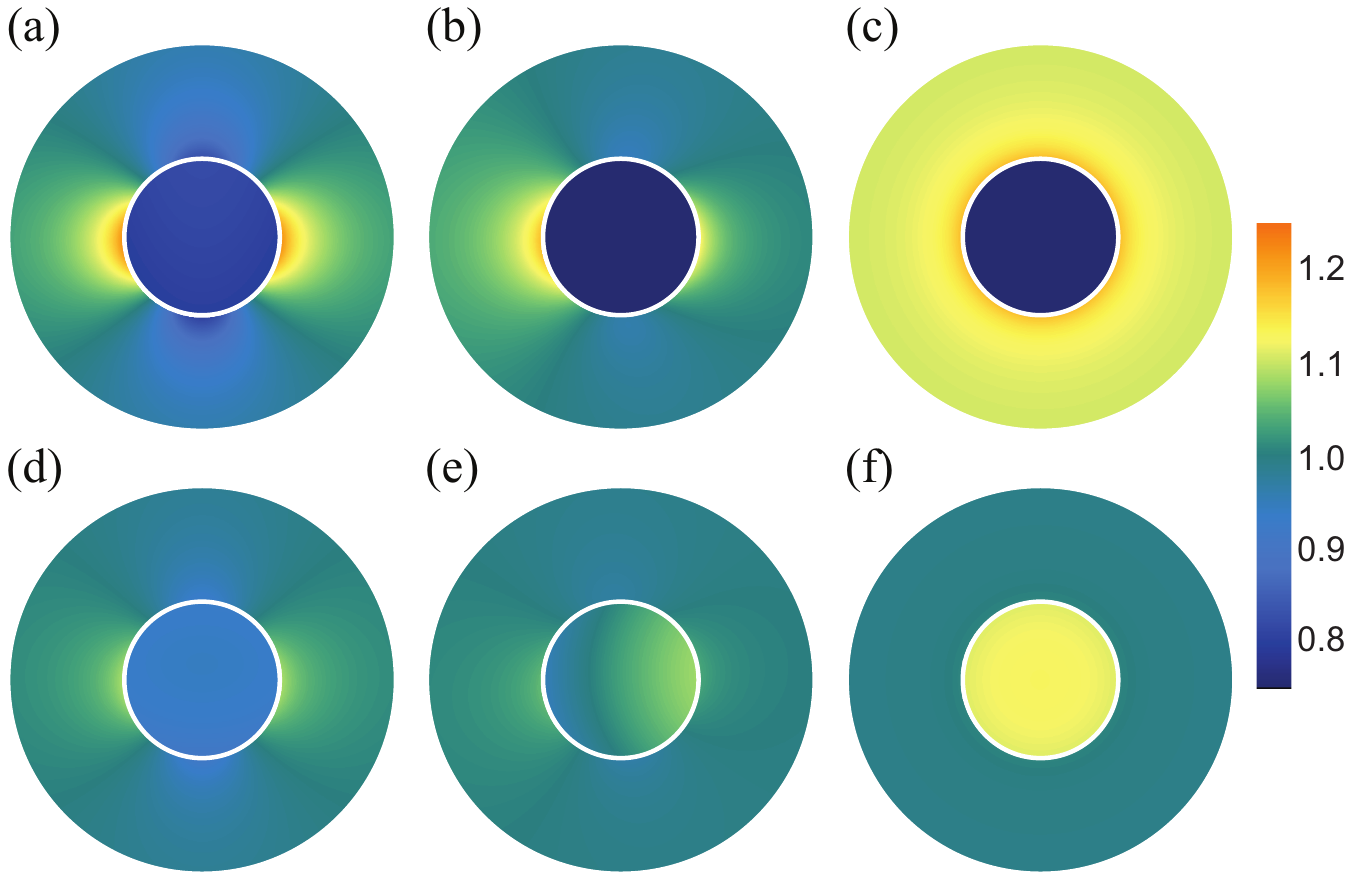}
\caption{Near-field distribution of amplitude of total electric field for a dielectric nanowire coated by graphene monolayer at (a) $\alpha=0^\circ$, (b) $\alpha=45^\circ$, and (c) $\alpha=90^\circ$ and helical graphene strips ($u=0.51$,  $\theta=51^\circ$) at (d) $\alpha=0^\circ$, (e) $\alpha=45^\circ$, and (f) $\alpha=90^\circ$. The nanowire is illuminated by dual-polarized wave with frequency of 13.15~THz. The surface of nanowire is depicted by the white line.}
\label{fig:fig9}
\end{figure}

Finally, the improved ability of helical graphene strips to suppress scattering of dual-polarized waves from dielectric nanowire at any incidence angle can be demonstrated by inspection of the near-field distribution for the amplitude $|{\bf E}|$ of total electric field. As desired cloaking frequency, we select 13.15~THz, which corresponds to the best performance of all-angle cloaking by graphene monolayer [see Fig.~\ref{fig:fig4}(a)]. This performance is then compared with that for helical graphene strips with optimal $u=0.51$ and $\theta=51^\circ$ (see Fig.~\ref{fig:fig8}). In our simulations, the amplitude of electric field is normalized to the averaged value $(2\pi)^{-1}\int_0^{2\pi}d\varphi|{\bf E}^i|$ of the amplitude of incident wave at $r=2.5R$. Therefore, the closer is the normalized amplitude of the total electric field outside the nanowire to unity, the more effective is scattering cancellation by the cloak. The near-field distribution of this amplitude for nanowires coated by graphene monolayer and helical graphene strips is shown in Fig.~~\ref{fig:fig9}. From this figure one can clearly see that helical graphene strips indeed possess improved ability for dual-polarized all-angle cloaking of dielectric nanowire, in agreement with simulated results for scattering efficiency.

\section{\label{concl}Conclusions}

Helical graphene strips have been proposed as a dual-polarized all-angle cloak for a dielectric nanowire. In the long-wavelength approximation, such strips has been treated as graphene-based metasurface with averaged conductivity of tensor form. Their cloaking performance in the terahertz band has been found to be better than that of graphene monolayer, even though the normally incident dual-polarized wave is considered. This is because helical graphene strips coiled around nanowire can effectively suppress scattering of both TE- and TM-polarized incident waves at single frequency due to cross-polarization coupling. It has been shown that this frequency can be widely tuned with pitch angle, period and width of graphene strips. Moreover, these parameters of the strips can be further optimized to ensure dual-polarized all-angle cloaking of dielectric nanowire at desired frequency. Such a property of helical graphene strips distinguishes them from other graphene-based metasurfaces used to cloak cylindrical objects in the terahertz band and makes them particular attractive for a number of applications, including noninvasive sensing and low-interference communication.

\section*{\label{ack}Acknowledgments}
This work was supported by Jilin University.

\section*{\label{app}Appendix A. Averaged boundary conditions}
\renewcommand{\theequation}{A\arabic{equation}}
\setcounter{equation}{0}

Following the approach of Ref.~\onlinecite{Shcherbinin_2018}, we derive here averaged boundary conditions at the surface of a dielectric cylinder coated by helical conducting strips. 

Consider a metasurface in the form of two-dimensional conducting strips placed periodically at the interface between dielectric cylinder of radius $R$ and outside ambient with relative constitutive parameters $\varepsilon_1$, $\mu_1$ and $\varepsilon_2$, $\mu_2$, respectively. The strips have the period $p$ and the width $w=up$. Let ${\bf i}_z$ and ${\bf i}_\varphi$ be the axial and the azimuthal unit vectors rotated by the angle $\theta$ with respect to the orthogonal vectors ${\bf i}_\parallel$ and ${\bf i}_\perp$, which are parallel and perpendicular to the strips, respectively. 

Consider an electromagnetic wave propagated inside both media. The wave field has the following form: $\{{\bf E},{\bf H}\}=\sum_{n=-\infty}^\infty \{{\bf E}^n,{\bf H}^n\}F_n$, where $F_n=\exp(-i\omega t+ ik_z z + in\varphi)$. Assume that the wavelength  $\lambda=2\pi\omega/c$ of the electromagnetic wave far exceeds the period $p$ of strips. In this case, there is a negligible contribution of the higher Bloch harmonics to the wave field at the metasurface area, where the averaged boundary conditions can be applied. For a dielectric cylinder coated by perfectly conducting strips, these conditions have the following form \cite{Bankov_1989}:
\begin{equation}
\begin{split}
& E_{\parallel 1} = E_{\parallel 2},~ E_{\perp 1} = E_{\perp 2}, \\
& E_{\parallel 1} = -iqZ_0N\mu_{\alpha}\ln\left[\sin^{-1}(0.5\pi u)\right]\left(H_{\perp 2}-H_{\perp 1}\right), \\
& H_{\parallel 2} - H_{\parallel 1} = iqZ_0^{-1}N\varepsilon_{\alpha}\ln\left[\sin^{-1}(0.5\pi(1- u))\right]E_{\perp 1},
\label{eq:app_a1} 
\end{split}
\end{equation}
where $q=1-k_\parallel^2/(\varepsilon_\alpha\mu_\alpha k^2)$, $N=2p/\lambda$, $\varepsilon_{\alpha} = \varepsilon_{1} + \varepsilon_{2}$, $\mu_{\alpha}=\mu_{1}\mu_{2}/(\mu_{1} + \mu_{2})$, $Z_0=\sqrt{\mu_0/\varepsilon_0}$ is the impedance of free space, $k_\parallel^2=\partial^2/\partial x_\parallel^2$ is the operator, $\partial/\partial x_\parallel= \sin\theta\partial/\partial z + \cos\theta R^{-1}\partial/\partial\varphi$, and $x_\parallel$ is the coordinate along the strips.

Averaged boundary conditions (\ref{eq:app_a1}) can be extended to the metasurface made of graphene strips with the finite conductivity $\sigma=\sigma(\omega)$. For this purpose, we introduce the graphene surface impedance $Z_g^{av}=u\sigma^{-1}$ averaged over the metasurface area. Using this impedance, we can rewrite the boundary conditions (\ref{eq:app_a1}) at $r=R$ as
\begin{equation}
\begin{split}
& E_{\parallel 1}^n = E_{\parallel 2}^n,~ H_{\parallel 2}^n - H_{\parallel 1}^n = -\sigma_\perp E_{\perp 1}^n, \\
& E_{\perp 1}^n = E_{\perp 2}^n,~ H_{\perp 2}^n - H_{\perp 1}^n = \sigma_\parallel E_{\parallel 1}^n,
\label{eq:app_a2} 
\end{split}
\end{equation}
where $\sigma_{\perp,\parallel} = (Z_{\perp,\parallel} + Z_g^{av})^{-1}$ are the principal components of the metasurface conductivity tensor
\begin{equation}
\hat {\boldsymbol \sigma} =\left( {\begin{matrix}
   {\sigma_\perp} & {0} \cr
   {0} & {\sigma_\parallel} \cr
\end{matrix}
} \right). \label{eq:app_a3}
\end{equation}
Here $Z_\parallel = -iqZ_0N\mu_{\alpha}\ln[\sin^{-1}(0.5\pi u)]$, $Z_\perp = iZ_0\left\{qN\varepsilon_{\alpha}\ln[\sin^{-1}(0.5\pi(1- u))]\right\}^{-1}$, and $k_\parallel^2=(k_z\sin\theta + nR^{-1}\cos\theta)^2$.

In the cylindrical coordinates $\{r,\varphi,z\}$, the conductivity tensor (\ref{eq:app_a3}) is non-diagonal and the boundary conditions (\ref{eq:app_a2}) are written as:
\begin{equation}
\begin{split}
& E_{z1} = E_{z2},~ H_{z2} - H_{z1} = -\sigma_{\varphi\varphi} E_{\varphi 1} - \sigma_{\varphi z} E_{z1}, \\
& E_{\varphi 1} = E_{\varphi 2},~ H_{\varphi 2} - H_{\varphi 1} =\sigma_{zz} E_{z 1} + \sigma_{z\varphi} E_{\varphi 1},
\label{eq:boundary_tens} 
\end{split}
\end{equation}
where $\sigma_{\varphi\varphi} = \sigma_\perp\cos^2\theta + \sigma_\parallel \sin^2\theta$, $\sigma_{zz} = \sigma_\parallel \cos^2\theta + \sigma_\perp\sin^2\theta$, and $\sigma_{\varphi z} = \sigma_{z\varphi} = (\sigma_\parallel - \sigma_\perp)\sin\theta\cos\theta$.

\section*{\label{app}Appendix B. Graphene description}
\renewcommand{\theequation}{B\arabic{equation}}
\setcounter{equation}{0}

Ignoring the quantum finite-size effect of graphene \cite{Abajo_2012}, the nanowire coating is treated as an infinitely thin graphene monolayer having the macroscopic surface conductivity $\sigma$, which depends on the angular frequency $\omega=2\pi f_r$, the chemical potential $\mu_c$, the ambient temperature $T$, and the scattering rate of charge carriers $\Gamma=1/\tau$. The surface conductivity of graphene consists of intraband and interband components $\sigma = \sigma_\textrm{intra} + \sigma_\textrm{inter}$, which are described by the Kubo formalism \cite{Falkovsky_2007}: 
\begin{widetext}
\begin{equation}
\begin{split}
& \sigma_\textrm{intra} = \frac{2 i e^2k_BT}{\hslash^2 \pi \left(\omega+i \Gamma\right)}\ln\left[2\cosh\left(\frac{\mu_c}{2k_BT} \right) \right],\\
&\sigma_\textrm{inter} = \frac{e^2}{4\hslash\pi}
\left[\frac{\pi}{2}+\arctan\left(\frac{\hslash\omega-2\mu_c}{2k_BT}\right) - \frac{i}{2}\ln\frac{\left(\hslash\omega+2\mu_c\right)^2}{\left(\hslash\omega-2\mu_c\right)^2+\left(2k_BT\right)^2} \right].
\end{split}
\label{eq:Kubo}  
\end{equation}
\end{widetext}
Here $k_B$ is the Boltzmann constant, $\hslash$ is the reduced Planck constant, and $e$ is the electron charge. The chemical potential $\mu_c$  is related to the carrier density $N_c$ as $\mu_c=\hslash v_F \sqrt{\pi N_c}$, where $v_F\simeq 10^6$~m/s is the Fermi velocity of electrons in graphene.


\bigskip

\bibliography{helical_graphene}

\end{document}